\documentclass[a4paper,11pt]{article}

\usepackage{amsfonts} 
\usepackage{amssymb} 
\usepackage{amsmath,amsthm} 
\usepackage{url}

\newtheorem{theoreme}{Theorem}
\newtheorem{lemma}[theoreme]{Lemma} 
\newtheorem{prop}[theoreme]{Proposition} 
 
\newtheorem{definition}[theoreme]{Definition}

\theoremstyle{remark}

\newtheorem{remark}[theoreme]{\bf Remark}

\newcommand{\be}{\begin{enumerate}}  \newcommand{\ee}{\end{enumerate}} 
\newcommand{\bi}{\begin{itemize}}  \newcommand{\ei}{\end{itemize}} 
\newcommand{\bd}{\begin{description}}  \newcommand{\ed}{\end{description}}

\newcommand{\comment}[1]{}

\def \CC {\mathbb{C}}
 
\def \NN {\mathbb{N}}

\def \DD {\mathcal{D}}

\DeclareMathOperator{\Tr}{Tr} 
\DeclareMathOperator{\vol}{vol} 
\DeclareMathOperator{\spec}{spec} 
\DeclareMathOperator{\diag}{diag} 
\DeclareMathOperator{\sh}{sinh} 
\DeclareMathOperator{\ch}{cosh}

\numberwithin{equation}{section}  
\renewcommand{\phi}{\varphi} 
\renewcommand{\epsilon}{\varepsilon} 
 
\title{\bf Resolvent expansions on hybrid manifolds}
\author{\Large Konstantin Pankrashkin \dag \and \Large Svetlana Roganova \ddag \and \Large Nader Yeganefar \S \\[\bigskipamount]
\dag{} Laboratoire de math\'ematiques d'Orsay, CNRS UMR 8628\\
Universit\'e Paris Sud XI, 91405 Orsay Cedex, France\\
E-mail \url{Konstantin.Pankrashkin@math.u-psud.fr}
\\[\medskipamount]
\ddag{} The work done at\\
Institut f\"ur Mathematik, Humboldt-Universit\"at\\
Rudower Chaussee 25, 12489 Berlin, Germany\\
Currently at the URSSAF Marseille, France\\
E-mail \url{roganova@gmail.com}\\[\medskipamount]
\S{} Laboratoire d'analyse, topologie, probabilit\'es, CNRS UMR 6632\\
Centre de Math\'ematiques et Informatique\\
39 rue Fr\'ed\'eric Joliot-Curie\\
13453 Marseille Cedex, France\\
E-mail \url{Nader.Yeganefar@cmi.univ-mrs.fr}\\[\bigskipamount]
\emph{Dedicated to the memory of Vladimir A. Geyler}
}

\date{}

\begin{document} 

\maketitle

\begin{abstract}
\noindent We study Laplace-type operators on hybrid manifolds, i.e. on configurations
consisting of closed two-dimensional manifolds and one-dimensional segments.
Such an operator can be constructed by using the Laplace-Beltrami operators on each component
with some boundary conditions at the points of gluing.
The large spectral parameter expansion of the trace of the second power of the resolvent
is obtained.
Some questions of the inverse spectral theory are adressed.
\end{abstract}



\section{Introduction}

The spectral theory on compact Riemannian
manifolds has been studied for a long time and takes its roots in physical problems. A great number of important results has been obtained and this subject has a lot of ramifications.  The main objects of investigation in spectral theory are Laplace type operators on a compact manifold, constructed from a Riemannian metric. A related subject which is developing very actively is the spectral theory on manifolds
with singularities, see e.g. \cite{brsing,cheeger}. Generalizing in another related direction, one can be interested
in the spectral theory on metric graphs, or quantum graphs \cite{AGA}, \ or more general multistructures \cite{mult}.
In the present paper we are studying some spectral properties of the Laplacian
on spaces composed of one- and two-dimensional pieces.

We consider the so-called ``hybrid manifolds'', which are unions of smoth manifolds connected by line segments.
More specifically, let us describe it from the topological point of view. Consider a set of $m$ compact Riemannian manifolds $M_1,...,M_m$ and 
a set of $n$ segments $L_1,...,L_n$. On each manifold $M_i$ we fix  some  points $q_{is}$, $s=1,\dots,\mu_i$, $\mu_i>0,\, i=1,\dots, m$. First of all we consider the disjoint union of all initial elements: $M_1\sqcup\dots\sqcup M_m\sqcup L_1\sqcup\dots\sqcup L_n$. 
Then we construct a one-to-one correspondence
between the set of end points of all segments and the set of points $q_{is}, s=1,\dots,\mu_i, i=1, \dots, M$. The following natural condition on the number of elements must be satisfied
\begin{gather*}
\sum_{i=1}^{m} \mu_i=2n.
\end{gather*}
Finally, according to this correspondence, we glue each end of each segment to
the corresponding point on one of the manifolds, which results in a certain topological space. 
We assume it to be path connected which immediately implies that $n\geqslant m-1$.
The topological space obtained by gluing the initial
manifolds and segments as described is called a \textit{hybrid manifold}.
If the manifolds are zero-dimensional, then we have a quantum graph. However, from now on,
we will assume that our manifolds are 2-dimensional. We will also often relabel the gluing points as $q_1, \ldots q_N$, where
\[
N=2n.
\]

Our first task will be to define Laplace type operators on a hybrid space. This construction involves the Laplacians on each smooth part
as well as boundary conditions at the points of gluing, and will be explained in Section 3. The spectral properties of the operators obtained in this way can be studied using
their resolvents or some function of it. In our approach we 
consider the trace of the squared resolvent. This is a standard procedure in spectral theory. Indeed, if we consider the usual (positive) Laplacian $\Delta$ on a Riemannian surface $M$, then it is known (see e.g. \cite{RS}) that  $R^2(z)=(\Delta+z^2)^{-2}$ is trace class.
If $G^2(x,y;z)$ is the integral kernel of this operator,
then we have the following  asymptotic expansion as $z\rightarrow \infty$ : for all $K\geqslant 0$,
\begin{gather}\label{4.3a}
G^2(x,x;z) = \sum_{k=0}^K \frac{a_k(x)\,\Gamma(k+1)}{4\pi z^{2k+2}}+O(z^{-2(K+2)}),
\end{gather}
where the coefficients $a_k$ are the local heat kernel invariants of $M$
and can be expressed by some universal expressions containing the metric tensor and
its derivates. Furthermore,
\begin{gather}\label{4.3}
\Tr R^2(z) = \sum_{k=0}^K \frac{a_k\,\Gamma(k+1)}{4\pi z^{2k+2}}+O(z^{-2(K+2)}),
\end{gather}
where the coefficients
\[
a_k=\int_M a_k(x)\,dx
\]
are the global heat kernel invariants of $M$. In particular, the first terms of this expansion reveal  important geometric and topological characteristics
of the surface:
\begin{gather} \label{trace}
\Tr R^2 \sim \frac{\vol M}{4\pi z^2}+\frac{\chi(M)}{6 z^4}+\cdots ,
\end{gather}
where $\vol M$ and $\chi(M)$ are the volume and the Euler characteristic of $M$, respectively;
this is closely related to the Weyl asymptotics for the eigenvalues, see e.g. \cite{brsing}.
Note that traces of some functions of the Laplacian are also of use
in the inverse spectral theory of quantum graphs, see e.g. \cite{jvb,GS,KN,roth}.

Our aim is to obtain similar results for a Laplace type operator on a hybrid manifold.
In fact, due to the singular structure of the hybrid space, we won't get an asymptotic expansion in the usual way
but in a certain extended form. More precisely, we show (subsection \ref{44})

\begin{theoreme}\label{intro1} Let $R(z)$ denote the resolvent of a Laplace type operator on a hybrid manifold. Then $\Tr R^2(z)$ has an expansion of the following form: for any $k\ge 4$ and $\alpha\in(0,1)$ one has 
\[
\Tr R^2(z) = \frac{\sum_{i} \vol M_i}{4\pi z^2}+ 
       \frac{\sum_{j}l_j}{4z^3}       
+\sum_{j=4}^k\frac{c_j(\ln z^2)}{z^k}+ O\Big(\frac{1}{z^{k+\alpha}}\Big),
\]
where $\vol M_i$ (resp. $l_j$) is the volume of the manifold $M_i$ (resp. of the segment $L_j$) used to construct the hybrif manifold, and the coefficients $c_j$ are rational functions.
\end{theoreme}
Each coefficient $c_j$ is actually a rational function of nonpositive degree. Therefore, each $c_j(\ln z^2)$ has an aymptotic expansion in powers of $1/\ln z^2$ for large $z$. For a rather large class of boundary conditions (see the beginning of Section 5 for more details), we are able to compute the first terms of the expansion of $c_k$. This allows us to address some quesions in inverse spectral theory : given the expansion of $\Tr R^2(z)$ in Theorem \ref{intro1}, can we recover some topological and metric parameters of the hybrid manifold from the coefficients? Under
some generic assumptions we prove the following result (subsection \ref{ssinv}):
\begin{theoreme}
Consider the expansion of the trace of the square of the resolvent as in Theorem \ref{intro1}. 
The knowledge of $Tr R^2$ determines:
\begin{itemize}
\item whether this manifold is hybrid or smooth;
\item the sum of the volumes of all manifolds taking part in the construction;
\item the sum of the Euler characteristics of all manifolds;
\item the number of segments used in this hybrid manifold;
\item the sum of the lengths of these segments;
\item the Euler characteristic of the hybrid manifold.
\end{itemize}
\end{theoreme}
We mention that a rather similar result for quantum graphs was obtained in \cite{PK}.

Finally, we will show in Theorem \ref{bord} that the expansion given in Theorem \ref{intro1} contains some information about the boundary conditions used to construct our Laplace type operator. We note that there is a certain canonical choice of boundary conditions
for quantum graphs resulting from considering shrinking manifolds approching the quantum graph \cite{EP};
no similar results are available for hybrid spaces, hence it is reasonable to take
into account a possibly large class of couplings. Our paper is based heavily
of the the PhD-thesis of the second named author \cite{disser} where
a class of the so-called disjoint boundary conditions was discussed.
Later in \cite{tolch} another class of boundary conditions was suggested.
We are treating here the most general boundary conditions. It comes out
that this requires a certain extension of the notion of pseudoasymptotic expansion introduced
in \cite{RS}, but the most essential features of the coefficients and their
interpretation are preserved.

\medskip

\emph{Acknowledgments.} As already mentioned, the paper is based and extends the PhD-thesis of Svetlana Roganova. She would like to thank her advisor Jochen Br\" uning for suggesting the problem and for many useful discussions, hints and comments.

\section{Definition of the Laplacian on a hybrid manifold}

\subsection{Basic facts on self-adjoint extensions}

We recall here some basic facts on the theory of self-adjoint
boundary conditions; for a more detailed presentation see e.g. \cite{bg,BGP}.

\begin{definition}\label{bvs}  Let $S$ be a closed densely defined
symmetric linear operator acting on a Hilbert space $\mathcal H$.
Let $\Gamma_1, \Gamma_2$ be two linear 
mappings from $\mathcal D(S^*)$ into a Hilbert space $\mathcal G$.
The triple 
$(\mathcal G,\Gamma_1, \Gamma_2)$ is called a
\textit{boundary value space} for $S$ if the following three conditions
are satisfied:
\begin{itemize}
\item for all $x, y \in \mathcal D(S^*)$ there holds
\[
\langle x,S^*y\rangle-\langle S^*x, y\rangle=\langle\Gamma_1x, \Gamma_2 y\rangle-\langle\Gamma_2 x, \Gamma_1 y\rangle,
\]
\item for any $u, v \in \mathcal G$ there exists $x \in \mathcal D(S^*)$ such that
$\Gamma_1 x=u$ and $\Gamma_2 x=v$,
\item $\ker \Gamma_1\cap\ker\Gamma_2=\DD(S)$.
\end{itemize} 
\end{definition}

The existence of a boundary triple in the above sense is equivalent to
the fact that the deficiency indices of $S$ are equal, e.g. that $S$
possesses self-adjoint extensions. Boundary triples deliver
a useful machinery for describing all self-adjoint extensions of $S$.

In what follows in this subsection  let $(\mathcal G,\Gamma_1,\Gamma_2)$ be a boundary triple
for $S$.

\begin{prop}\label{prop-bvs}
(1) Let $A$ and $B$ be bounded linear operators in $\mathcal G$
satisfying
\[
AB^*=BA^*\text{ and }
\ker\begin{pmatrix}
A & -B\\
B & A
\end{pmatrix}=0,
\]
then the restriction $H^{A,B}$ of $S^*$ to the vectors $f\in\DD(S^*)$
with $A\Gamma_1 f=B\Gamma_2 f$ is a self-adjoint extension of $S$.
Moreover, each self-adjoint extension of $S$ can be represented in such a form.

(2) There exists a bijection between unitary operators $U$ in $\mathcal G$
and self-adjoint extensions $H^U$ fo $S$ given by
$U\leftrightarrow H^U=H^{A,B}$ with $A=1-U$ and $B=i(1+U)$.

\end{prop}

Due to the above proposition the operator $H_0$ which is the restriction of
$S^*$ to $\ker\Gamma_1$ is self-adjoint. For $z\notin\spec H_0$ the restriction of $\Gamma_1$ onto $\ker(S^*-z)$ is a bijection;
denote the inverse operator by $\gamma(z)$. Clearly, $\gamma(z)$ acts from
$\mathcal G$ to $\mathcal H$, and we will refer to this operator as to \emph{$\gamma$-field}
for the boundary triple, and the operator $Q(z):=\Gamma_2\gamma(z)$ will be called
\emph{the Weyl function}.

\begin{prop}[Krein resolvent formula]\label{kh}
For $z\notin\spec H^{A,B}\cup\spec H_0$ there holds
\begin{equation}
(H^{A,B}-z)^{-1}=(H_0-z)^{-1}-\gamma(z)\big(BQ(z)-A\big)^{-1}B\gamma(\bar z)^*. \label{formula}
\end{equation}
\end{prop}

\subsection{Example 1: Laplacian on a segment}\label{ssex1}

In the Hilbert space $\mathcal H=L^2[0,l]$, $l>0$, consider
the operator $S=-d^2/dt^2$ with the domain
$\DD(S)=\{f\in H^2[0,l]: f(0)=f(l)=f'(0)=f'(l))\}$.
One can easily show that the adjoint operator $S^*$
is given by the same expression on the domain
$\DD(S^*)=H^2[0,l]$. By direct calculation one can verify
that as a boundary triple one can take
\[
\mathcal G=\CC^2,
\quad
\Gamma_1 f=\begin{pmatrix}
-f'(0)\\f'(l)
\end{pmatrix},
\quad
\Gamma_2 f=\begin{pmatrix}
f(0)\\f (l)
\end{pmatrix}.
\]
The distinguished self-adjoint extension $H_0$ of $S$, i.e. the restriction of $S^*$
to $\ker \Gamma_1$ is exactly the operator $-d^2/dx^2$ with the Neumann boundary conditions.

To construct the corresponding $\gamma$-field $\gamma(z)$ and the Weyl function $Q(z)$ it is useful to use
the Green function $G_0(x,y;z)$, i.e. the integral kernel of the resolvent
$(H_0-z)^{-1}$, $z\notin \spec H_0$, which is given by
\begin{equation}\label{neum}
G_0(x,y;z)=\begin{cases}
\dfrac{\ch \big(\sqrt{-z}(x-l)\big)\ch \big(\sqrt{-z}y\big)}{\sqrt{-z}\sh\big(\sqrt{-z}l\big)},& \quad x\geq y,\\[\bigskipamount]
\dfrac{\ch \big(\sqrt{-z}(y-l)\big)\ch \big(\sqrt{-z}x\big)}{\sqrt{-z}\sh\big(\sqrt{-z}l\big)}
,& \quad x\leq y,
\end{cases}
\end{equation}
where the branch of the square root ix fixed by $\Im \sqrt{\lambda}\ge 0$.
In terms of $G_0$ one has
\[
\gamma(z) \begin{pmatrix}
\xi_1\\ \xi_2
\end{pmatrix}=
\xi_1 G_0(\cdot,0;z)
+\xi_2 G_0(\cdot,l;z)
\]
and
\begin{multline*}
Q(z)=\begin{pmatrix}
G_0(0,0;z) & G_0(l,0;z)\\
G_0(0,l;z) & G_0(l,l;z)
\end{pmatrix}={}\\
\dfrac{1}{\sqrt{-z}\sh l\sqrt{-z}}
\begin{pmatrix}
\ch l\sqrt{-z} & 1\\
1 & \ch l\sqrt{-z}
\end{pmatrix}.
\end{multline*}

\subsection{Example 2: Laplacian on a manifold}\label{ssex2}

Let $M$ be a two-dimensional closed manifold and $\Delta=\Delta_M$ be the closure of the positive
Laplace-Beltrami operator defined initially on $C^\infty(M)$.
It is known that $\DD(\Delta)\subset C^0(M)$. Fix some (mutually distinct) points $q_1,\dots, q_\mu\in M$
and denote by $S$ the restriction of $\Delta$ to the functions vanishing at all these points
(which makes sense, as all functions from the domain are continuous).
The defficiency indices of $S$ are $(\mu,\mu)$.

Let $G(x,y;z)$ be the Green function of $\Delta$, i.e. the integral kernel
of the resolvent $(\Delta-z)^{-1}$, then, for $z\notin\spec \Delta$, 
there holds
\[
\DD(S^*)=\big\{
f=f_0+\sum_{j=1}^\mu \varphi_j G(\cdot, q_j;z):\,
f_0\in\DD(\Delta), \quad \Phi=(\varphi_1,\dots,\varphi_\mu)\in\CC^\mu
\big\}.
\]
The Green function can be presented as follows:
\[
G(x,y;z)=\dfrac{1}{2\pi} \ln\dfrac{1}{d(x,y)}+F(x,y;z),
\]
where the second term $F$ is continuous in $M\times M$.
Using the above decompositions one can show that there exist
uniquely defined linear functionals $a_j$, $b_j$ on $\DD(S^*)$, $j=1,\dots,\mu$, such that
\[
f(x)=\dfrac{1}{2\pi}\ln\dfrac{1}{d(x,y)}\, a_j(f) +b_j(f) + o(1)
\text{ for } x\to q_j,
\]
and as a boundary triple for $S$ one can take
$(\CC^\mu,\Gamma_1,\Gamma_2)$ with
\[
\Gamma_1 f=\big(a_j(f)\big)_{j=1,\dots,\mu},
\quad
\Gamma_2 f=\big(b_j(f)\big)_{j=1,\dots,\mu}.
\]
Note that the distinguished extension of $S$ corresponding to the boundary condition $\Gamma_1f=0$
is just the initial operator $\Delta$.
By direct calculation one can show that the $\gamma$-field $\gamma(z)$
and the Weyl function $Q(z)$ for the above boundary triple are given by
\[
\gamma(z)(\xi_j)=\sum_j \xi_j G(\cdot,q_j;z), \quad
Q_{jk}(z)=\begin{cases}
G(q_j,q_k;z),& j\ne k,\\
F(q_j,q_j;z),& j=k.
\end{cases}
\]
More details to the above constructions can be found in \cite{BGP}.
We also mention the paper \cite{LH} treating a trace formula for the operators of the above type.

Below we use the following asymptotic expransion, see \cite[Formula (38)]{A}:
for all $M\geqslant 1$,
\begin{equation}
 \label{eq-fxx}
F(x,x,z)= \frac{1}{4\pi}\left(-\ln z^2-2\gamma+
\sum_{n=1}^M \frac{\Gamma(n) a_{n}(x,x)}{z^{2n}}\right)+O(z^{-2(M+1)}).
\end{equation}
Here $a_{n}(x,x)$ is the $n$-th local heat kernel coefficient and $\gamma$ is Euler's constant.

\subsection{Hamiltonian on a hybrid manifold}\label{ssdef}

Using the formalism of the previous section, we are going to construct Laplace type operators on hybrid manifolds and study some of their properties. 

Consider a hybrid manifold constructed with $m$ compact manifolds $M_1, \ldots , M_m$,
with marked points $q_{is}\in M_i$, $s=1, \ldots , \mu _i$, and $n$ segments $L_1, \ldots , L_n$, as described in the introduction. On each $M_i$ consider also the usual Laplace-Beltrami operator $\Delta_{M_i}$, which is a self-adjoint operator on $L^2(M_i)$, with domain $\mathcal{D}(\Delta_{M_i})$, 
the second Sobolev space of $M_i$.  Denote by $D_i$ restriction of $\Delta_{M_i}$ to the domain
\begin{gather}\label{D_i}
\mathcal{D}(D_i)=\{f\in \mathcal{D}(\Delta_{M_i}): f(q_{is})=0, \quad s=1,\ldots ,\mu _i\}.
\end{gather}

We parametrize the segments $L_j$ by $\{x_j\in\mathbb{R}: x_j\in [0, l_j]\}$, where $j=1,\dots,n$ and denote  by $D^s_j$ the closure in $L^2(L_j)$ of the operator $-\frac{d^2}{dx_j^2}$ defined
on $C_0^\infty(L_j)$. 

The operator $S$ defined by 
\begin{gather}
S=D_1\oplus\dots\oplus D_m\oplus D^s_1\oplus\dots\oplus D^s_n \label{D}
\end{gather} 
is a symmetric operator in $L^2(M_1)\oplus\dots\oplus L^2(M_m)\oplus L^2(L_j)\oplus \dots\oplus L^2(L_n)$ with deficiency indices $(4n, 4n)$.
\begin{definition} \label{s}
A Laplace type operator on a hybrid manifold is a self-adjoint extension of the operator $S$.
\end{definition}
We use the constructions of the previous section to describe all
self-adjoint extentions of $S$.
Namely, for each operator $D_j$ take the boundary triple $\big(\CC^{\mu_j},\Gamma_{1,M_j},\Gamma_{2,M_j}\big)$
defined in subsection~\ref{ssex2} and for each operator $D^s_j$ take the boundary triple $\big(\CC^2,\Gamma_{1,L_j},\Gamma_{2,L_j}\big)$ as defined in subsection~\ref{ssex1}. Then, obviously, their direct sum $(\CC^{2N},\Gamma_1,\Gamma_2)$, where
$\Gamma_j=\Gamma_{j,M}\oplus\Gamma_{j,L}$, and 
$\Gamma_{j,M}=\Gamma_{j,M_1}\oplus\Gamma_{j,M_m}$,
$\Gamma_{j,L}=\Gamma_{j,L_1}\oplus\dots\oplus \Gamma_{j,L_n}$,
$j=1,2$, is a boundary triple for $S$.
The corresponding $\gamma$-field $\gamma(z)$ and the Weyl function are also the direct sums
of the corresponding objects of each part.

Below we use the enumeration of the points $\{q_{is}\}=:\{q_j\}$
in such a way that the point $q_j$ is the one attached to the initial point
of $L_{k}$ if $j=2k-1$, $k\in\NN$, or the the terminal vertex of $L_k$ if $j=2k$, $k\in\NN$.

Now we are going to restrict the class of admissible boundary conditions.
It is natural to suppose that the boundary coniditions at each point of gluing
are local, i.e. involves only the boundary values of the functions at this point.
In view of the above enumeration of the points $q_j$ this means that
we consider only the boundary conditions of the form $A\Gamma_1=B\Gamma_2$
where each of the $2N\times 2N$ matrices $A$ and $B$ consists of four diagonal blocks
\[
A=\begin{pmatrix}
(a_{i,i}) & (a_{i,i+N})\\
(a_{i+N,i}) & (a_{i+N,i+N})
\end{pmatrix},
\quad
B=\begin{pmatrix}
(b_{i,i}) & (b_{i,i+N})\\
(b_{i+N,i}) & (b_{i+N,i+N})
\end{pmatrix}.
\]
It is useful to introduce the $2\times 2$ matrices
\[
A_j=\begin{pmatrix}
a_{i,i} & a_{i,i+N}\\
a_{i+N,i} & a_{i+N,i+N}
\end{pmatrix},
\quad
B_j=\begin{pmatrix}
b_{i,i} & b_{i,i+N}\\
b_{i+N,i} & b_{i+N,i+N}
\end{pmatrix},
\]
clearly, these matrices should satisfy the same conditions $A_j B_j^*=AB_j^*$
and $\det(A_jA_j^*+B_jB_j^*)\ne 0$ for each $j$.
It is also clear, by proposition \ref{prop-bvs}, that each of the boundary conditions can be rewritten
using unitary matrices of the same structure
\[
A_j=1-\Phi_j, \quad B_j=i(1+\Phi_j)
\]
with
\[
\Phi_i=\begin{pmatrix}
\phi_{i,i} & \phi_{i,i+N}\\
\phi_{i+N,i} & \phi_{i+N,i+N}
\end{pmatrix}\in U(2)
\]
and
\[
\Phi=\begin{pmatrix}
(\phi_{i,i}) & (\phi_{i,i+N})\\
(\phi_{i+N,i}) & (\phi_{i+N,i+N})
\end{pmatrix}\in U(2N).
\]

This paramatrization will be referred to as \emph{canonical}.
Note that diagonal matrices $\Phi_j$ correspond to the absence of coupling
at $q_j$: the boundary conditions split into one boundary condition on the manifold
and another one on the segment attached; we will call such boundary conditions \emph{reducible}.
Below we consider non-reducible boundary conditions only, hence assuming that all
the matrices $\Phi_j$ in the above constructions are non-diagonal.

\section{The resolvent expansion}\label{developpement}

{\bf Warning.} \emph{From now on, due to technical reasons
and in order to be compatible with other works,
we will use $-z^2$ as our spectral parameter, and not $z$. However, we will still use e.g. the notation $R_0(z)$ to denote $(S_0+z^2)^{-1}$, and not $(S_0-z)^{-1}$ as in the previous section.} 
Note for further use that in this notation the Green function on manifolds or segments obeys
\[
\int_M G(x,u,z)G(u,y,z)\,du=-\frac{1}{2z}\,G'_z(x,y,z)
\]
(which is just a consequence of the Hilbert resolvent identity).

\subsection{Computation of $\Tr R^2$}  

Let us start with the computation of $ \Tr R^2$ using Theorem \ref{kh}. Setting
\[
T(z)=\gamma(z)[BQ(z)-A]^{-1}B(\gamma(\bar z))^*,
\] 
we obtain 
 \begin{gather*} 
\Tr R^2(z)=\Tr R_0^2(z)-\Tr(R_0(z)T(z))-\Tr(T(z)R_0(z))+\Tr T^2(z).     
\end{gather*}

We recall that the resolvent $R_0$ is the direct sum of the 
resolvents of the ordinary Laplacians on the manifolds $M_i$ (denoted by $R_{0, M_i}$) and the resolvents for the Neumann Laplacians on the segments $L_j$ (denoted by $R_{0, L_j}$)  forming the hybrid manifold.
For the manifolds one has 
\begin{multline*}
\Tr R_{0, M_i}^2(z)= \int_{M_i}\int_{M_i}
G_{M_i}(x,u,z)G_{M_i}(u,x,z) \, dx\, du\\
=-\int_{M_i}\frac{1}{2z}(G_{M_i})'_z(x,x,z)\, dx,
\end{multline*}
and for the segments one uses the same formula with the explicit expression \eqref{neum};
we will use it later.

The linear operator $T(z)$ can be also rewritten as an integral operator as follows.
Denote the entries of the $2N\times 2N$ matrix $[BQ(z)-A]^{-1} B$ by $c_{ij}(z)$. Using 
$G(x,y,z)=\overline{G(y,x,\bar{z})}$ we get, for any $f\in L^2$:
\begin{gather*}
T(z)f(y)=\gamma(z)[BQ(z)-A]^{-1}B(\gamma(\bar z))^*f(y)=\\
\sum_{i,j}\int c_{ij}(z) G(y,q_i,z)f(u)G(q_j,u,z)\, du.
\end{gather*}
The integral here is in fact the sum of the integrals over manifolds and 
segments. Using this integral representation of the operator $A$, we calculate 
the remaining terms of $\Tr R^2$ as follows. First, for any $f\in L^2$:
\begin{gather*}
R_0(z)T(z)f(x)=\sum_{i,j}c_{ij}(z)\int\int G(x,y,z)G(y,q_i,z)f(u)G(q_j,u,z) \,dy \,du.
\end{gather*}
Hence, the operator $R_0(z)T(z)$ has the integral kernel
\begin{gather*}
K(x,t,z)=\sum_{i,j}c_{ij}(z)\int G(x,y,z)G(y,q_i,z)G(q_j,t,z)\, dy,
\end{gather*}
therefore 
\begin{gather*}
\Tr R_0(z)T(z)=\sum_{i,j}c_{ij}(z)\int\int G(x,y,z)G(y,q_i,z)G(q_j,x,z)\, dx\, dy,
\end{gather*}
or, 
\begin{multline*}
\Tr R_0(z)T(z)=
 \frac{1}{2}\sum_{i,j}c_{ij}(z)\int \left(\int G(x,y,z)G(y,q_i,z)\, dy
  \right)G(q_j,x,z)\, dx \\
  +\frac{1}{2}\sum_{i,j}c_{ij}(z)\int \left(\int G(x,y,z)G(q_j,x,z)\, dx
   \right)G(y,q_i,z)\, dy\\
   =\frac{1}{2}\sum_{i,j}c_{ij}(z)\Big(\int-\frac{1}{2z}G'_z(x,q_i,z)
   G(q_j,x,z)\, dx+\\
   \int-\frac{1}{2z}G'_z(q_j,y,z)G(y,q_i,z)\, dy\Big)\\
   =-\frac{1}{4z}\sum_{i,j}c_{ij}(z)
   \int \left(G(q_j,x,z)G'_z(x,q_i,z)+G'_z(q_j,x,z)G(x,q_i,z)\right)\, dx\\
   =-\frac{1}{4z}\sum_{i,j}c_{ij}(z)\int\left(G(q_j,x,z)G(x,q_i,z)\right)'_z\, dx\\
   =\frac{1}{8z}\sum_{i,j}c_{ij}(z)
   \left(\frac{1}{z}G'_z(q_j,q_i,z)\right)'_z\\
   =-\sum_{i,j}c_{ij}(z)\left(\frac{G'_z(q_j,q_i,z)}{8z^3}-\frac{G''_{zz}(q_j,q_i,z)}
   {8z^2}\right).
\end{multline*}

Let us now find $\Tr T^2(z)$. We will use the same method and represent the operator
$T^2(z)$ as an integral operator:
\begin{multline*}
T^2(z)f=\sum_{i,j}c_{ij}(z)\int G(x,q_i,z)G(q_j,y,z) T(z)f(y)\, du\\
=\sum_{i,j,k,l} c_{ij}(z)\, c_{kl}(z)\int G(x,q_i,z)G(q_j,y,z)\times{}\\
\bigg(\int G(y,q_k,z)G(q_l,u,z) f(u)\, du\bigg)
\, dy\\
=\sum_{i,j,k,l} c_{ij}(z)\, c_{kl}(z)\int G(x,q_i,z)G(y,q_k,z)G(q_j,y,z)G(q_l,u,z)f(u)\, du\, dy.
\end{multline*}

Now we find:
\begin{multline*}
\Tr T^2(z)\\
=\sum_{i,j,k,l} c_{ij}(z)\, c_{kl}(z)\int G(x,q_i,z)G(q_l,x,z)G(y,q_k,z)G(q_j,y,z)\, dy\, dx\\
=\frac{1}{4z^2}\sum_{i,j,k,l}c_{ij}(z)\, c_{kl}(z)G'_z(q_l,q_i,z) G'_z(q_j,q_k,z).
\end{multline*}

Summarizing the results obtained, we state
\begin{theoreme} \label{tr1}Consider a hybrid manifold $H$, consisting of the manifolds $M_i$ and the segments $L_j$,
and the Laplace operator corresponding to the boundary conditions $A\Gamma_1=B\Gamma_2$ on $H$.
Then the following formula for the trace of the square of the resolvent of this operator holds:
\begin{multline*}
\Tr R^2(z)= -\int_H\frac{1}{2z}G'_z(x,x,z)\, dx
\\
+2\sum_{i,j=1}^{2N}c_{ij}(z)\left(\frac{G'_z(q_j,q_i,z)}{8z^3}
-\frac{G''_{zz}(q_j,q_i,z)}
   {8z^2}\right)\\
   +\frac{1}{4z^2}\sum_{i,j,k,l=1}^{2N}c_{ij}(z)\, c_{kl}(z)
   G'_z(q_l,q_i,z) G'_z(q_j,q_k,z).
\end{multline*}
Here $G(x,y,z)$ is the Green function of $S_0$ on the hybrid manifold and the entries of the matrix $[BQ(z)-A]^{-1} B$ are denoted by $c_{ij}(z)_{1\leqslant i,j\leqslant 2N}$.
\end{theoreme}

\subsection{The asymptotic expansion}
The result of Theorem \ref{tr1} is given in terms of the Green functions for the Laplacians 
on the smooth parts of the hybrid manifold. But using it in 
applications is practically impossible for two reasons: first of all, inverting the $2N \times 2N$ matrix $BQ(z)-A$ can be difficult, and secondly, an explicit computation of the Green function is almost never possible.
 Nevertheless, there are ways to get some simplifications if we restrict attention to large $z$. As we will see below, 
 the special  structure of the matrix $(BQ(z)-A)$ allows us to find its inverse asymptotically
 as $z\rightarrow \infty$. Likewise,  one can use the representation of $R_0$ as an asymptotic series in powers of $z$ 
 (analog of heat kernel expansion).  Thus we may attempt to generalize \eqref{trace}
to the case of hybrid manifolds.

 For further calculations the following lemma is crucial. 
\begin{lemma} \label{2} Let $M$ be a compact two-dimensional manifold 
and $G(x,y;z)$ be the integral kernel of $(\Delta+z^2)^{-1}$.
For any $x,y\in M$, $x\ne y$, and any $k=0,1,2,\dots$,
there exists $C_k=C_{k,x,y}>0$ such that for $z\to+\infty$ one has
\[
\dfrac{\partial^k G(x,y,z)}{\partial z^k}=O(e^{-C_kz}).
\]
\end{lemma}
\begin{proof} Let $P(x,y;t)$ be the heat kernel of $\Delta$,
i.e. the integral kernel for the semigroup $e^{-t\Delta}$.
It is known \cite[Corollary 6]{dav} that there are positive constants $a$
and $\delta$ such that
\[
0\le P(x,y;t)\le \dfrac{a}{t}\, \exp\Big(
-\dfrac{d(x,y)^2}{4(1+\delta) t}.
\Big)
\]
Clearly,
\[
G(x,y,z)=\int_0^{+\infty} e^{-z^2 t} P(x,y;t)dt,
\]
hence
\[
\dfrac{\partial^k G(x,y,z)}{\partial z^k}=
\int_0^{+\infty} p_k(z,t) e^{-z^2 t} P(x,y;t)dt,
\]
where $p_k$ is a certain polynomial. One can estimate
\[
\Big|\dfrac{\partial^k G(x,y,z)}{\partial z^k}\Big|\le 
a \int_0^{+\infty} \dfrac{q_k(z,t)}{t}
\exp\Big( -\dfrac{d(x,y)^2}{4(1+\delta) t}\Big)
e^{-z^2 t} \,dt,
\]
where $q_k$ is another polynomial. Now let us note the identity
\[
\int_0^{+\infty} t^{\nu-1} e^{-b/t} e^{- pt}dt
=2 \big(b/p)^{\nu/2} K_\nu(2\sqrt{bp}),
\]
where $K_\nu$ is the modified Bessel function,
see \cite[Supplement 4.3]{polyanin}, which gives, with some
polynomials $q_j$,
\[
\Big|\dfrac{\partial^k G(x,y,z)}{\partial z^k}\Big|\le 
2a \sum_{j=0}^{K} q_j(z) 
\Big(\dfrac{d(x,y)^2}{4(1+\delta)}\Big)^{j/2}
K_j\Big(\dfrac{zd(x,y)}{\sqrt{1+\delta}}
\Big).
\]
Now it is sufficient to recall the asymptotic behavior
$K_\nu(s)=O(e^{-s})$ for $s\to+\infty$.
\end {proof}

Thus, 
if the points $x$ and $y$ do not coincide, the Green function $G(x,y,z)$, as well as its derivatives $G'_z(x,y,z)$ and $G''_{zz}(x,y,z)$
decay exponentially as $z$ tends to infinity. 

In the statement of Theorem \ref {tr1} the terms $G(q_i,q_j,z)$ appear, where the distance 
between the points $q_j$ and $q_j$ is fixed by the configuration of our hybrid space. 
Hence all terms of such type for non-coinciding points $q_i\neq q_j$ are exponentially small as $z\rightarrow\infty$, so we need to take into account only the diagonal terms $G(q_i,q_i,z)$.
Using this observation and performing  calculations similar to the proof 
of Theorem  \ref {tr1}, but neglecting  terms of exponential small order
of $z$ we prove that for some positive constant $c$, as $z\rightarrow\infty$ 
\begin{multline*}\label{tr2}
\Tr R^2(z)= -\int_H\frac{1}{2z}G'_z(x,x,z)\, dx\\
+2\sum_{i=1}^{2N}c_{ii}(z)\left(\frac{G'_z(q_i,q_i,z)}{8z^3}-\frac{G''_{zz}(q_i,q_i,z)}
   {8z^2}\right)\\
   +\frac{1}{4z^2}\sum_{i,j=1}^{2N}c_{ij}(z)\, c_{ji}(z)
   G'_z(q_i,q_i,z) G'_z(q_j,q_j,z)+ O(e^{-cz}).
\end{multline*}
This result can be rewritten in a compact matrix-form, as we are going to see now. Let us
denote by $G'_z$ the diagonal matrix with entries $G'_z(q_i,q_i,z)$. Denote also by $G''_{zz}$ the diagonal matrix with entries $G''_{zz}(q_i,q_i,z)$. This is a slight abuse of notation, since $G'_z$ and $G''_{zz}$ are also well defined operators; it will be clear from the context what is meant by this notation.
Then it is easy to see that \eqref{tr2} can be formulated as follows.

\begin{theoreme} \label{tr3}We have
the following 
asymptotic relation as $z\rightarrow\infty$:
\begin{gather*}
\Tr R^2(z)= -\int_{H} \frac{G'_z(x,x,z)}{2z}\, dx -\frac{1}{4z^2}\Tr\left(G''_{zz}\,[BQ-A]^{-1}B\right)\\
+\frac{1}{4z^3} \Tr\left(G'_z\,[BQ-A]^{-1}B\right)+\frac{1}{4z^2} 
\Tr \left(G'_z\,[BQ-A]^{-1}B\right) ^2+ O(e^{-cz}).
\end{gather*}
\end{theoreme}

Now we need only to compute  $(BQ(z)-A)^{-1}B$, and fortunately, 
this is possible in this approximation.

As we have already mentioned, the matrix $Q(z)$ has a direct product representation,
$Q(z)=Q_1\oplus\dots Q_m\oplus G$ where $Q_i$ is the $Q$-matrix for the $i$-th manifold $M_i$ in the hybrid manifold
(see subsection \ref{ssex2}), for example,
\begin{align*}
     Q_1=\begin{pmatrix}
     F(q_1,q_1,z)&...&G(q_1,q_{\mu_1},z) \\
     ...&...&...\\
     G(q_{\mu_1},q_1,z)&...&F(q_{\mu_1},q_{\mu_1},z)
     \end{pmatrix}
\end{align*}
and $G=(G_{kl})$ is an $N\times N$-matrix consisting of the Green functions (denoted by $G_s$) for the Neumann Laplacian on the segments:
\begin{align*}
G_{kl}=\begin{cases}
  G_s(q_k,q_l,z) &\text{if $q_k$ and $q_l$ belong to the same segment,}\\
  0& \text{otherwise.}\\
  \end{cases}
\end{align*}

Nevertheless, the matrix $B Q(z)-A$ is still too complicated to find its inverse explicitly. 
But letting $z\rightarrow \infty$ and applying Lemma \ref{2} the situation improves radically. More precisely, all non-diagonal terms of $Q(z)$ are either zero or of type $G(q_i, q_j,z), \, i\neq j$, and decay exponentially as 
$z$ tends to infinity. Therefore, we can write $Q(z)=Q_d(z)+\tilde{Q}(z)$, where $\tilde{Q}(z)$ is exponentially 
small as $z$ goes to infinity, and  the leading order approximation $Q_d(z)$ to $Q(z)$ is given by
\[
Q_d(z)=\diag\Big(F(q_1,q_1,z), \dots,F(q_N,q_N,z),G_s(q_1,q_1,z),\dots,G_s(q_N,q_N,z)\Big).
\]
Then $(B\big(Q(z)-A)^{-1}B$ to first order, i.e. $(B Q_d(z)-A)^{-1}$, consists of four diagonal 
$N\times N$ blocks. The inverse matrix can be found using the Frobenius formula for
a block matrix consisting of matrices $A,B,C,D$:
\begin{align*}
\begin{pmatrix}
    A&B\\
    C&D\\
\end{pmatrix}^{-1} = 
\begin{pmatrix}
    (A-BD^{-1}C)^{-1}&A^{-1}B(CA^{-1}B-D)^{-1}\\
    (CA^{-1}B-D)^{-1}CA^{-1}&(D-CA^{-1}B)^{-1}\\
\end{pmatrix}.
\end{align*}
Hence one can obtain the block representation
\begin{align*}
[B Q_d(z)-A]^{-1} B = 
\begin{pmatrix}
    X^{-1} U& X^{-1}W\\
    X^{-1} \overline{ W\rule{0ex}{2ex}}& X^{-1}V\\
\end{pmatrix},
\end{align*}
where  $U$, $W$, $V$, $X$ are diagonal $N\times N$ matrices with
\begin{multline}
 \label{eq-matx}
X_{i,i}\equiv X_i=(b_{i,i} b_{i+N,i+N}-b_{i,i+N}b_{i+N,i})F_i G_i\\
+ (a_{i+N,i}b_{i,i+N}-a_{i,i}b_{i+N,i+N})G_i + (a_{i,i+N} b_{i+N,i} - a_{i+N,i+N}b_{i,i})F_i\\
+(a_{i,i}a_{i+N,i+N}-a_{i,i+N}a_{i+N,i}),
\end{multline}
and
\begin{equation}
       \label{eq-uv}
\begin{aligned}
U_{i,i}\equiv U_i&=\big( b_{i,i} b_{i+N,i+N}-b_{i,i+N}b_{i+N,i}\big) G_i +(a_{i,i+N}b_{i+N,i} - a_{i+N,i+N} b_{i,i}),\\
W_{i,i}\equiv W_i&=a_{i,i+N}b_{i+N,i+N} - a_{i+N,i+N} b_{i,i+N},\\
V_{i,i}\equiv V_i&=\big( b_{i,i} b_{i+N,i+N}-b_{i,i+N}b_{i+N,i}\big) F_i + (a_{i+N,i}b_{i,i+N}-a_{i,i}b_{i+N,i+N}).
\end{aligned}
\end{equation}
Note that if we change the parameters $A\to L A$, $B\to LB$, with a non-degenrate matrix $L$
of the same structure as $A$ or $B$, then all of the above expressions acquire the factor $\det L_j$.

It is important to mention that the matrix $X_i$ is invertible for large $z$. To see this,
let us note first that all four summands in \eqref{eq-matx} have different asymptotics for large $z$:
by \eqref{eq-fxx} one has $F_i\sim \ln z^2$, and by direct calculation,
\[
G_i(z)=1/z+O(e^{-cz}),
\]
hence it is sufficient to show that at least one of them is non-zero.
To show this, it is useful to assume that the matrices $A$ and $B$ are taken in the canonical form,
see subsection \ref{ssdef}.
In this case, assuming that all four summands in \eqref{eq-matx}
vanish one arrives for each $i$ to the incompatible system
\begin{gather}
1+\phi_{i,i}+\phi_{i+N,i+N}+\det \Phi_i=0, \quad 1-\phi_{i,i}-\phi_{i+N,i+N}+\det \Phi_i=0,\\
1-\phi_{i,i}+\phi_{i+N,i+N}-\det \Phi_i=0, \quad 1+\phi_{i,i}-\phi_{i+N,i+N}-\det \Phi_i=0.
\end{gather}

\begin{lemma} \label{estimate} With the notation above, we have
\[
[BQ(z)-A]^{-1}=[B Q_d(z)-A]^{-1}+O(e^{-cz}),
\]
for some positive constant $c$.
\end{lemma}
\begin{proof}
First, we write
\begin{align*}
[BQ(z)-A]^{-1} &= \Big[B\big(Q_d(z)+\tilde{Q}(z)\big)-A\Big]^{-1}\\
                  &= [B Q_d(z)-A]^{-1}[1+B \tilde Q(z)[B Q_d(z)-A]^{-1}]^{-1}.
\end{align*}
As can be easily seen from the expressions for  $U_{ii}$, $W_{ii}$ and $V_{ii}$, one has
$[B Q_d(z)-A]^{-1}B=O(z^2)$. As  $\tilde{Q}(z)$ decays exponentially fast for large $z$, the product  $[BQ_d(z)-A]^{-1}\tilde{Q}(z)$ will then
 also be exponentially small. By using the Neumann series to compute $[1+[BQ_d(z)-A]^{-1}\tilde{Q}(z)]^{-1}$,
we see that this term is of the form $1+O(e^{-cz})$, and this immediately implies the conclusion of the Lemma.

\end{proof}

Lemma \ref{estimate} allows us to reformulate theorem \ref{tr3} as follows:

\begin{theoreme}\label{th-expan}  For $z\to+\infty$ there holds, up to  $O(e^{-cz})$ terms,
\begin{multline}
         \label{eq-trz0}
\Tr R^2(z)= -\int_{H} \frac{G'_z(x,x,z)}{2z}\, dx
-\dfrac{1}{4z^2} \sum_i \dfrac{F''_i U_i+G''_i V_i}{X_i}\\
+\dfrac{1}{4z^3} \sum_i \dfrac{F'_i U_i+G'_i V_i}{X_i}
+\dfrac{1}{4z^2} \sum_i \dfrac{(F'_i U_i)^2+2F'_iG '_i |W_i|^2+(G'_i V_i)^2}{X_i^2}.
\end{multline}
\end{theoreme}

This theorem describes the large spectral parameter asymptotic of the trace of the second power of the resolvent for a Laplace operator through the Green functions of the manifolds and segments
forming the hybrid manifold. In the next section, we will write this expansion  in terms of heat kernel coefficients for the manifolds forming the hybrid manifold. 

\subsection{Pseudoasymptotic resolvent expansion in terms of heat kernel coefficients} \label{44}

To compute the above expression for the trace in terms of heat kernel coefficients, 
we need to consider the term 
\[
\int_{H} \frac{G'_z(x,x,z)}{2z}\, dx.
\]
In fact, this term consists of two types of integrals: over the manifolds 
and over the segments.
As we already know (see equation \eqref{4.3}), for manifolds the following expansion holds as $z\rightarrow \infty$: for all $q\geqslant 0$
\[
-\int_{M} \frac{G'_z(x,x,z)}{2z}\, dx=\Tr R_0^2(z)
=\sum_{k=0}^{q} \frac{a_k\,\Gamma(k+1)}{4\pi z^{2k+2}}+O(z^{-2(q+2)}),
\]
where $R_0$ is the resolvent of the Laplace operator on this manifold and the constants $a_k$ are the
heat kernel coefficients for it. 
The integral over segments can be easily computed asymptotically using the 
explicit expressions \eqref{neum}:
\begin{gather*}
-\int_0^l \frac{G'_z(x,x,z)}{2z}\, dx=\frac{l}{4z^3}+\frac{1}{2z^4}+O(e^{-cz}),
\quad z\to +\infty.
\end{gather*}

\begin{definition}\label{devz}
Assume that there exist rational functions $c_k$, $k=1,2,\dots$, such that
for any $q\ge 0$ and $\alpha\in(0,1)$ a function $f=f(z)$ admits the following representation for $z\to+\infty$:
\begin{equation}
       \label{eq-fzexp}
f(z)=\sum_{k=0}^{q} \frac{c_{k}(\ln z^2)}{z^k}+O\left(\frac{1}{z^{q+\alpha}}\right).
\end{equation}
An expansion of the form \eqref{eq-fzexp} will be called \emph{pseudoasymptotic},
and in this case we write
\[
f(z)\sim \sum_{k=0}^{\infty} \frac{c_{k}(\ln z^2)}{z^k}.
\]
\end{definition}

\begin{lemma}\label{dev}
If a function $f$ admits a pseudoasymptotic expansion, then
this expansion is unique.
\end{lemma}

\begin{proof}
First, we notice that if $P$ and $Q$ are two polynomials
and $\alpha>0$, then, for large $z$,
\[
z^{-\alpha}=o\Big(
\dfrac {P(\ln z^2)}{Q(\ln z^2)}
\Big).
\]
Suppose that there are two pseudoasymptotic expansions of $f$, i.e. for any $q\ge 0$ and any $\alpha>0$
\begin{equation}
    \label{eq-fekv}
f(z)=\sum_{k=0}^{q} \frac{c_{k}(\ln z^2)}{z^k}+O\left(\frac{1}{z^{q+\alpha}}\right)=
\sum_{k=0}^{q} \frac{d_{k}(\ln z^2)}{z^k}+O\left(\frac{1}{z^{q+\alpha}}\right).
\end{equation}
For $q=0$ we obtain
\[
c_0(\ln z^2)-d_0(\ln z^2)=O(z^{-\alpha})
\]
which implies $c_0=d_0$. 
By an immediate induction on $k$, we show similarly that we will always obtain $c_k(\ln z^2)-d_k(\ln z^2)=O(z^{-\alpha})$
and consequently the uniqueness  of the expansion.
\end{proof}

\begin{remark} Let us notice that if a function $f$ possesses a pseudoasymptotic expansion, and if  
each coefficient $c_k$ is a rational function with nonpositive degree, then each $c_k$
 can be uniquely expanded as a $\ln z^2$-asymptotic series: for all $K\geq 0$,
\[
c_k(\ln z^2)=\sum_{l=0}^{K}\frac{c_{kl}}{(\ln z^2)^l}+O\left(\frac{1}{(\ln z^2)^{K+1}}\right),
\] for some constants $c_{kl}$. 
\end{remark}

\begin{theoreme} \label{12} The trace $\Tr R(z)^2$ obtained in Theorem \ref{th-expan}
admits a pseudoasymptotic expansion of the form 
\begin{gather*}
\Tr R^2(z) = \frac{\sum_{i} \vol(M_i)}{4\pi z^2}+ 
       \frac{\sum_{j}l_j}{4z^3}+
\sum_{k\ge 4} \dfrac{c_k(\ln z^2)}{z^k},
\end{gather*}
where
$c_4(\lambda)=\dfrac{\sum_j \chi(M_j)}{6}+\dfrac{N}{4}
+\dfrac{N_0}{\lambda}+o\Big(
\dfrac{1}{\lambda}
\Big), \quad \lambda\to+\infty
$
and
\[
N_0=\#\big\{j\in\{1,\dots, N\}: a_{i,i+N}b_{i+N,i}-a_{i+N,i+N}b_{i,i}\ne 0\big\}.
\]
Moreover, if $N_0=N$, then $c_k=O(1)$ as $z\to+\infty$ for all $k$.
\end{theoreme}

\begin{proof}
Consider first more carefully
the expression \eqref{eq-matx} for $X_i$.
Let us show first that at least one of the last two terms on the right-hand side,
$\alpha_j F_i$ with $\alpha_j=a_{i,i+N} b_{i+N,i} - a_{i+N,i+N}b_{i,i}$
and $\beta_j:=a_{i,i}a_{i+N,i+N}-a_{i,i+N}a_{i+N,i}\equiv \det A_j$
is non-vanishing. Taking the matrices $A_j$ and $B_j$ in the canonical form
$A_j=1-U_j$, $B_j=i(1+U_j)$, and assuming that
the two  above terms vanish one arrives at the system
\[
1+\phi_{i,i}-\phi_{i+N,i+N}-\det \Phi_j=0,
\quad
1-\phi_{i,i}-\phi_{i+N,i+N}+\det \Phi_j=0,
\]
and one immediately obtains $\varphi_{i+N,i+N}=1$,
which means that the matrix $\Phi_j$ is diagonal.
As this case is excluded by assumption, we have the result.
Hence on has always $1/X_i=O(1)$ for large $z$.

Consider the second sum on the right-hand side of \eqref{eq-trz0}; the other terms
are considered in the same way. Each summand is of the form
\[
E_j:=\dfrac{F''_i U_i+G''_i V_i}{X_i}
\] 

Assume first that $a_{i,i+N}b_{i+N,i}-a_{i+N,i+N}b_{i,i}\ne 0$. In this case, substituting
the expansion \eqref{eq-fxx} for $F_i$ into the expression in Theorem \ref{th-expan}
one arrives at an expansion of the form
\begin{multline*}
E_j=
\dfrac{\dfrac{\ln z^2}{z^5}+ \sum_{k=0}^K\dfrac{d'_k}{z^k}+ O(\dfrac{1}{z^{K+1}})}{d''_0+\ln z^2+ \dfrac{\ln z^2}{z}+ \sum_{k=1}^K\dfrac{d''_k}{z^k}+O(\frac{1}{z^{K+1}})}\\
=\dfrac{\dfrac{\ln z^2}{z^5}+ \sum_{k=0}^K\dfrac{d'_k}{z^k}+ O(\dfrac{1}{z^{K+1}})}{\ln(z^2e^{d''_0})\left(1+\dfrac{\ln z^2}{z\ln(z^2e^{d''_0})}+ \sum_{k=1}^K\dfrac{d''_k}{z^k\ln(z^2e^{d''_0})}+O(\dfrac{1}{z^{K+1}})\right)}.
\end{multline*}
For large $z$ the last term in the denominator can be expanded as 
\begin{gather*}
\left(1+\frac{\ln z^2}{z\ln(z^2e^{d''_0})}+ \sum_{k=1}^K\frac{d''_k}{z^k\ln(z^2e^{d''_0})}+O(\frac{1}{z^{K+1}})\right)^{-1}\\
=1+\sum_{l=1}^{L} \frac{g_l\ln^{s_1}(z^2)}{\ln^{s_2}(z^2e^{d''_0})z^{l}}+O\left(\frac{1}{z^{L+1}}\right),
\end{gather*}
for some constant $g_l$ and non-negative powers $s_1, s_2$. On can see also that $s_1\leqslant s_2$. Multiplying both series in powers of $z$ in the expression for the first fraction  we obtain an expansion of type
\begin{gather*}
\sum_{l=0}^{L} \frac{g'_l\ln^{t_1}(z^2)}{\ln^{t_2}(z^2e^{d''_0})z^{l}}+O\left(\frac{1}{z^{L+1}}\right)=\sum_{l=0}^{L} \frac{c_{l}(\ln z^2)}{z^l}+O\left(\frac{1}{z^{L+1}}\right),
\end{gather*}
with coefficients 
\[
c_{l}(\ln z^2)=\frac{P_l(\ln z^2)}{Q_l(\ln z^2)}
\]
which are rational in $\ln z^2$. Note that in this first case we always have
$\deg P_l\le \deg Q_l$.

The case $a_{i,i+N}b_{i+N,i}-a_{i+N,i+N}b_{i,i}=0$ can be treated in a similar way,
one deals with an expansion of the form
\[
E_j\sim \dfrac{\dfrac{\ln z^2}{z^5}+ \sum_{k=0}^K\dfrac{d'_k}{z^k}+ O(\dfrac{1}{z^{K+1}})}{1+\dfrac{\ln z^2}{z}+ \sum_{k=1}^K\dfrac{d''_k}{z^k}+O(\frac{1}{z^{K+1}})}
\]
and develops the denominator directly using the geometric series.

Furthermore, examining the asymptotic behavior of all the functions,
one can easily conclude that  the last three sums on the right-hand sum
of \eqref{eq-trz0} are of order $O(z^{-4})$, and the only terms that
contribute to the coefficient $c_4$ come from the summands
\[
K_j:=-\dfrac{F''_i U_i}{ 4z^2 X_i}+\dfrac{F'_i U_i}{4z^3 X_i}.
\]
Note that if $a_{i,i+N}b_{i+N,i}-a_{i+N,i+N}b_{i,i}=0$, then
$U_i=O(z^{-1})$ and $K_j$ does not contribute to $c_4$.
Otherwise, by considering the leading terms in the numerators
and the denominators,
\[
K_j=\dfrac{1+o(1)}{\ln z^2}.
\]
Summing over all $j$ we obtain the result.

\end{proof}

\begin{remark}
It is useful to emphasize that all the ``coefficients'' $c_k$ are non-constant, i.e.,
we have always some logarithmic terms. Indeed, it is clear that
the logarithmic terms can only disappear if the coefficients
of the terms containing $F_i$ in the expressions \eqref{eq-matx}
and \eqref{eq-uv} vanish.

Taking the matrices $A$ and $B$ in the canonical form (subsection \ref{ssdef}),
this condition would mean that
\[
1+\phi_{i,i}+\phi_{i+N,i+N}+\det \Phi_j=0, \quad
1+\phi_{i,i}-\phi_{i+N,i+N}-\det \Phi_j=0.
\]
This implies directly $\varphi_{i,i}=-1$, and the matrix $\Phi_j$
must be hence diagonal, which contradicts to the non-reducibility
of the boundary conditions.

\end{remark}

\begin{remark}
The expansion obtained allows one to address some questions of the inverse spectral theory.
It is clear that the knowledge of the coefficients in this expansion gives
the summary volume of the manifolds and the summary length of
the one-dimensional parts. Considering the fourth term one can obtain some estimates
about the Euler characteristics of the manifolds and the number of the segments.
Note that for generic boundary conditions one has $N_0=N$
hence generically one can recover the sum of the Euler characteristics
of the manifolds and the number of the segments. We will adress such questions in greater details
in the next section.
\end{remark}

Let us first note  that the formula for $\Tr R^2$  obtained
in Theorem \ref{12} depends on the heat kernel coefficients
for the smooth parts of the hybrid manifold, and it is well known that heat kernel coefficients are recursive.
Some recursivity of the coefficients in the expansion of $\Tr R^2$ follows as well.
But as a series inversion is required, we cannot obtain this recursive 
formula in an explicit way. Nevertheless, it is possible to find
the formula for some terms in this expansion for some special class of boundary conditions. 

\section{Special case of boundary conditions}

As can be seen from the proof of theorem \ref{12}, the boundary
conditions enter the pseudoasymptotic expansion in a rather involved way.
In this section we treat in greater details the case when
the matrix $B$ is invertible. In this case one can normalize
by $B=1$, and $A$ must be then self-adjoint with the same
block structre as previously. To distinguish from the previous
case on the notation level, we denote the matrix elements of $A$
by $\lambda$,
\[
   A= \begin{pmatrix}
    (\lambda_{i,i})&(\lambda_{i, i+N})\\
    (\overline{\lambda_{i,i+N}})&(\lambda_{i+N, i+N})\\
     \end{pmatrix},
\]
and the diagonal entries are real. We assume additionally that
the entries $\lambda_{i+N,i+N}$ are non-zero for all $i\in\{1,\dots, N\}$.

Note that the expressions \eqref{eq-matx} and \eqref{eq-uv} simplify,
\begin{gather*}
X_{i,}=(F_i -\lambda_{i,i})(G_i-\lambda_{i+N,i+N})-|\lambda_{i,i+N}|^2,\\
U_{i}=G_i - \lambda_{i+N,i+N},\quad
W_{i}=\lambda_{i,i+N}, \quad V_{i}=F_i -\lambda_{i,i}.
\end{gather*}
Theorem \ref{12} is, of course, applicable, and one has additionally $N_0=N$.
As noted in the proof of Theorem \ref{12} we also have $c(\lambda)=O(1)$
for large $\lambda$.

To simplify notation in this section, we will use the following convention.
If the function $f=f(z,\ln{z^2})$ has a $z$-pseudoasymptotic expansion with coefficients $c_q(\ln{z^2})$, we noticed before the statement of Theorem \ref{12} that each $c_q(\ln{z^2})$ can be expanded as
\[
c_q(\ln z^2)=\sum_{k=0}^{K}\frac{c_{qk}}{(\ln z^2)^k}+O\left(\frac{1}{(\ln z^2)^{K+1}}\right),
\]
for some constants $c_{qk}$. We will then write
\[
f\sim \sum_{q,k=0}^{\infty}\frac{c_{qk}}{z^q(\ln{z^2})^k}.
\]

\subsection{First terms in the resolvent expansion.}

\begin{lemma} \label{5.1} Let $n\geq 4$.
The coefficient $c_n$ appearing in Theorem \ref{12} has an expansion
in $1/log z^2$, and the first term in this expansion is of the following form:
\begin{gather*}
\frac{\sum_{i}\chi(M_i)}{6}+\frac{N}{4}, \quad n=4;\\
\sum_{i=1}^N \frac{2k-1}{4\lambda _{i+N,i+N}^{2k+1}},\quad n=2k+1, \,k>1;\\
\sum_{i}\frac{a_{ki}\Gamma(k+1)}{4\pi}+\sum^{N}_{i=1}  \frac{2k}{4\lambda _{i+N,i+N}^{2k-2}}, \quad n=2k+2, \,k>1.
\end{gather*}
\end{lemma}

 \begin{proof}
 
The terms in $c_4$ and the first term in $c_n$ for even $n$ arise from
 the expansions of the second power of the resolvents
for all manifolds and segments. The other terms require some calculation. 
Let us look once more at the expression in Theorem \ref{th-expan} and analyze 
it more carefully. As we can see 
the expansion of 
\[
\frac{1}{(F_i-\lambda_1)(G_i -\lambda_2)-|\lambda_3|^2)}
\] 
always has $\ln z^2$ in the denominator, i.e. for some non-zero constants $k_{nm}''$ 
one has the $z$-pseudoasymptotic expansion
\begin{gather*}
\dfrac{1}{D_i}
\sim \frac{4\pi}{\lambda_{i+N,i+N}\ln z^2}+\sum_{n,m=0}^\infty\dfrac{k''_{nm}}
{z^n(\ln z^2)^{m+1}},
\end{gather*}
where we denote $(F_i-\lambda_{i,i})(\frac{1}{z} -\lambda_{i+N,i+N})-|\lambda_{i,i+N}|^2)$ by $D_i$.
In the corresponding numerators we have terms of type $F_i''$, $F_i'$ and $F_i$. 
Their expansions as $z\rightarrow \infty$ are: for all $p\geqslant 1$
\begin{gather*}
F_i(x,x,z)=\frac{1}{4\pi}\left(-2\gamma-\ln z^2+
\sum_{n=1}^p \frac{\Gamma(n) a_{ni}(x,x)}{z^{2n}}\right)+O(z^{-2p-2)}),\\
F'_i(x,x,z)=\frac{-1}{2\pi z}-\sum_{n=1}^p \frac{2n\Gamma(n) a_{ni}(x,x)}{4\pi z^{2n+1}}+O(z^{-2p-3)}),\\
F''_i(x,x,z)=\frac{1}{2\pi z^2}+\sum_{n=1}^p \frac{2n(2n+1)\Gamma(n) a_{ni}(x,x)}{4\pi z^{2n+2}}+O(z^{-2p-4}).
\end{gather*}
It is clear now that the first terms in the expansion of $c_n$ (i.e. terms without $\ln z^2$) can appear only in terms containing $F_i$. All other terms will contain this logarithm in denominator. Moreover, only the part $\frac{-\ln z^2}{4\pi}$ in $F_i$ matters.

The same arguments are valid for the last summand whose denominator contains
$(\ln z^2)^2$, and consequently we should take only the term $(F_i-\lambda_{i,i})^2$ 
into account.
Finally, we try to find the terms containing only powers of $z$ in 
\begin{gather*}
\frac{-2F_i-F_i}{4z^5D_i}
+\frac{(F_i-\lambda_{i,i})^2}{4z^6(D_i)^2}.
\end{gather*}
First of all let us treat the denominator more carefully
\begin{gather*}
\dfrac{1}{D_i}
\sim \frac{4 \pi}{\lambda_{i+N,i+N}\ln z^2}\left(1-\frac{1}{z\lambda_{i+N,i+N}}+
\sum_{n=0}^\infty\dfrac{k'_n}{z^n\ln z^2}\right)^{-1}\\
\sim \frac{4 \pi}{\lambda_{i+N,i+N}\ln z^2}\left(1+\sum_{n=1}^\infty\dfrac{1}{(z\lambda_{i+N,i+N})^n}+
\sum_{n,m=0}^\infty\dfrac{k''_n}{z^n(\ln z^2)^{m+1}}\right),
\end{gather*}
with some coefficients $k_n', k_n''$, whose explicit expressions are not important now.

Then we see that in $\frac{-3F_i}{4z^5D_i}$ the terms containing only powers of $z$
(without logarithm) are
\begin{gather*}
\frac{-3}{4z^5}\cdot \frac{4\pi}{\lambda_{i+N,i+N}\ln z^2}
\left(1+\sum_{n=1}^\infty\dfrac{1}{(z\lambda_{i+N,i+N})^n}\right)\cdot \frac{-\ln z^2}{4\pi}\\=
\frac{3}{4} \sum_{n=0}^\infty \dfrac{1}{z^{n+5}\lambda_{i+N,i+N}^{n+1}}.
\end{gather*}

Performing the same reasoning we find that
\begin{gather*}
\dfrac{1}{D_i^2}\sim \frac{(4\pi)^2}{(\lambda_{i+N,i+N}\ln z^2)^2}\left(
\sum_{n=0}^\infty \dfrac{n+1}{z^n\lambda_{i+N,i+N}^n} +
\sum_{n,m=0}^\infty\dfrac{k''_n}{z^n(\ln z^2)^{m+1}}\right).
\end{gather*}
The only term which cancels this logarithm in $(F_i -\lambda_{i,i})^2$ is
$\frac{(\ln z^2)^2}{(4\pi)^2}$. And the contribution to the "pure"\, polynomial part in $z$
 is 
\begin{gather*}
\frac{1}{4z^6}\cdot\frac{(4\pi)^2}{(\lambda_{i+N,i+N}\ln z^2)^2}\sum_{n=0}^\infty \dfrac{n+1}{z^n\lambda_{i+N,i+N}^n}\cdot\frac{(\ln z^2)^2}{(4\pi)^2}=\frac{1}{4}\sum_{n=0}^\infty \frac{n+1}{z^{n+6}\lambda_{i+N,i+N}^{n+2}}.
\end{gather*}

Finally, we find
(not taking into account those which arise from the expansion of $R_0$)

\begin{gather*}
\frac{3}{4} \sum_{n=0}^\infty \dfrac{1}{z^{n+5}\lambda_{i+N,i+N}^{n+1}}+
\frac{1}{4}\sum_{n=0}^\infty \frac{n+1}{z^{n+6}\lambda_{i+N,i+N}^{n+2}}=
\frac{1}{4}\sum_{n=0}^\infty \frac{n+3}{z^{n+5}\lambda_{i+N,i+N}^{n+1}}.
\end{gather*}
Collecting the terms with the same power of $z$ we prove the lemma.
\end{proof}

The next result requires more complicated calculations of the same nature, 
so we just state the main steps of its proof.

{\lemma \label{5.2}  Fix some integer $n\geq 4$ and consider the coefficient $c_n$ which  appears in Theorem \ref{12}. The second term of the expansion of $c_n$ 
has the following form:
\begin{gather*}
\sum^{N}_{i=1} \frac{\pi (n-4)(n-2)|\lambda_{i, i+N}|^2}{\lambda^{n-3}_{i+N, i+N}\ln z^2}, \quad n=2k+1, \, k>1;\\
\sum^{N}_{i=1} \frac{\pi (n-4)(n-2)|\lambda_{i, i+N}|^2}{\lambda^{n-3}_{i+N, i+N}\ln z^2}+\sum^{N}_{i=1} \frac{a_{li}(l+1)!}{\ln z^2}, \quad n=2l+4, \, l\geqslant 0.
\end{gather*}

}

\begin{proof}
The idea is the same as before: one can obtain the terms of this form only from specific terms in the  expansion. 
Due to the special structure of the denominator we have terms of type $\frac{f(z)}{\ln z^2}$ and terms $\frac{g(z)\ln z^2}{\ln^2 z^2}$ and the fact that 
in the numerator only the terms with $F_i$ contain a logarithm, allows us to restrict our calculation to some specific terms.

We need as before the expansion of $\frac{1}{D_i}$ but up to the second power of $\ln z^2$
and the expansion of$\frac{1}{D^2_i}$ up to the third power. 
To simplify the expressions we denote
the following quantity by $W_i$
\begin{gather*}
W_i=-2\gamma-4\pi\left(\lambda_{i,i}-\frac{|\lambda_{i,i+N}|^2}{\lambda_{i+N,i+N}}\right)+
\frac{2\gamma+4\pi\lambda_{i,i}}{z\lambda_{i+N,i+N}}\\
+\left(1-\frac{1}{z\lambda_{i+N,i+N}}\right)\sum_{n=1}^{\infty}\frac{(n-1)!a_{ni}}{z^{2n}}
\end{gather*}
and state that up to the second power of $\ln z^2$
\begin{gather*}
\frac{1}{D_i}\sim \frac{4\pi}{\lambda_{i+N,i+N}\ln z^2}\sum_{n=0}^{\infty}\left(\frac{1}{z\lambda_{i+N,i+N}}\right)^n\\
+\frac{4\pi W_i}{\lambda_{i+N,i+N}\ln^2 z^2}\sum_{n=1}^{\infty}n\left(\frac{1}{z\lambda_{i+N,i+N}}\right)^{n-1}
\end{gather*}

and up to the third power
\begin{gather*}
\frac{1}{D^2_i}\sim \frac{16\pi^2}{\lambda^2_{i+N,i+N}\ln^2 z^2}\sum_{n=0}^{\infty}\frac{n}{(z\lambda_{i+N,i+N})^{n-1}}\\
+\frac{4\pi^2 W_i}{\lambda^2_{i+N,i+N}\ln^3 z^2}\sum_{n=0}^{\infty}\frac{n(n-1)}{(z\lambda_{i+N,i+N})^{n-2}}.
\end{gather*}

Then the terms we are looking for appear in
\begin{gather*}
\frac{1}{D_i}\left(\frac{-F''_i(\frac{1}{z}-\lambda_{i+N,i+N})}{4z^2}-\frac{3(F_i-\lambda_{i,i})}{4z^5}+\frac{F'_i(\frac{1}{z}-\lambda_{i+N,i+N})}{4z^3}\right),
\end{gather*}
and in 
\begin{gather*}
\frac{1}{D^2_i}\left(\frac{(F_i-\lambda_{i,i})^2}{4z^6}\right).
\end{gather*}

Expanding and summing these expressions we find that the terms containing the first
power of logarithm in the denominator can be arranged in two sums as in the statement of the Lemma.
\end{proof}

In the same way, but using much more complicated calculations, we can get
the following result whose proof we omit (see \cite{disser} for details).
{\lemma \label{5.3}In Theorem \ref{12} the terms with $\ln^2 z^2$ 
in the denominator of the $c_n$ are (we do not separate the different powers of $z$
in order to not complicate the expression):
\begin{gather*}
\frac{1}{\ln^2 z^2}\sum^N_{i=1}\Bigg(
\frac{1}{z^4}-\frac{2\gamma}{z^4}-\frac{4\pi\lambda_{2i}}{z^4}+
\sum^\infty_{n=1}\frac{(2n+1)(n-1)!a_{ni}}{z^{2n+4}}+(2\pi\lambda_{2i}\\-\gamma)\frac{2\pi\lambda_{3i}^2}{z^4\lambda_{1i}}\sum^\infty_{n=1}\frac{n(n+2)}{(z\lambda_{1i})^n}
-\gamma\sum^\infty_{n=1}\frac{2(n+1)n!a_{ni}}{z^{2n+4}}\\+\frac{2\pi^2\lambda_{3i}^4}{\lambda_{1i}^2 z^4}\sum^\infty_{n=1}\frac{n(n^2-1)}{(z\lambda_{1i})^{n-1}}
-4\pi\lambda_{2i}\sum^\infty_{n=1}\frac{(n+1)n!a_{ni}}{z^{2n+4}}
+\frac{4\pi\lambda_{3i}^2}{z^4\lambda_{1i}}\sum^\infty_{n=0}\frac{n+1}{(z\lambda_{1i})^{n}}
\\+\sum^\infty_{n=1}\frac{(n-1)!a_{ni}}{z^{2n}}\sum^\infty_{k=1}\frac{(k+1)k!a_{ki}}{z^{2k+4}}+
\frac{\pi\lambda_{3i}^2}{z^4\lambda_{1i}}\bigg(\sum^\infty_{n=1}\frac{(n-1)!a_{ni}}{z^{2n}} \sum^\infty_{k=1}\frac{k(k+2)}{(z\lambda_{1i})^{k}}
\\+\sum^\infty_{n=1}\frac{2(2n+1)n!a_{ni}}{z^{2n}}\sum^\infty_{k=0}\frac{1}{(z\lambda_{1i})^{k}}
+\sum^\infty_{n=1}\frac{2n!a_{ni}}{z^{2n}}\sum^\infty_{k=0}\frac{1+2k}{(z\lambda_{1i})^k}\bigg)\Bigg),
\end{gather*}
where $\lambda_{1i}=\lambda_{i+N,i+N}$, $\lambda_{2i}=\lambda_{i,i}$ and 
$\lambda_{3i}=|\lambda_{i,i+N}|$.}

\subsection{Inverse spectral data}\label{ssinv}

Being restricted to the class of the boundary conditions introduced in the beginning of the section,
we consider now the inverse spectral problem, i.e. the question ``Which information 
about the initial system can one obtain using the expansion of the second power of the resolvent?''.
As we will see, it is possible to find some geometric characteristics of 
the system and some information about the operator on it.

\begin{theoreme} Consider the expansion of the trace of the square of the resolvent as in the Theorem \ref{12}. 
The knowledge of $Tr R^2$ determines:
\begin{itemize}
\item whether this manifold is hybrid or smooth;
\item the sum of the volumes of all manifolds taking part in the construction;
\item the sum of the Euler characteristics of all manifolds;
\item the number of segments used in this hybrid manifold;
\item the sum of the lengths of these segments;
\item the Euler characteristic of the hybrid manifold.
\end{itemize}
\end{theoreme}
\begin{proof}
The presence of $\ln z^2$-type terms is a criteria of singularity. 
If in the expansion of $Tr R^2$ there are no such terms, this means that the 
considered manifold is a "normal"\, manifold without any singular points. Indeed, the
$\log$-terms appear only from the singularities of the Green function
(and not its derivatives) at the points of gluing.

The coefficient of $z^{-2}$ is equal to $\sum \vol M_i/(4\pi)$ and provides us with the sum of the volumes of all manifolds.

The coefficient of $z^{-3}$ is equal to $\sum  l_j/4$ and provides us with the sum of the lengths of all segments.

Considering the term of type $1/(z^4\ln z^2)$ we find the number $N/2$ of all segments.

The coefficient of $z^{-4}$ is equal to 
\[
\dfrac{\sum \chi(M_i)}{6}+\dfrac{N}{4}
\]
and gives as the sum of the Euler characteristics of all manifolds since
we already know the number $N$.

Finally, using a Mayer-Vietoris type argument, see \cite{disser} for details,
it is not difficult to show that the Euler characteristic of the hybrid manifold is $\sum \chi(M_i)-N$, which is easy to find now.
\end{proof}

The results obtained in this work do not let us find the volume of each manifold 
and the length of each segment separately. But it should be possible to find these 
quantities with the use of scattering theory \cite{KN}.\\
The resolvent expansion provides us also with some information about how we glue the segments to the manifold in the hybrid space. Namely, we can 
obtain some information about the matrix of boundary conditions which 
defines the Laplace operator on the hybrid space.

\begin{theoreme}\label{bord} Consider the $z$-pseudoasymptotic expansion of the trace of the square of the resolvent expansion of type \ref{12}. If we assume that we know the heat kernel coefficients for all manifolds composing the hybrid manifold, and that the coefficients $\lambda_{i+N, i+N}$ are mutually distinct and nonzero,
we can find the diagonal elements of the matrix of boundary conditions 
$A$ and the absolute values of its non-diagonal elements up to  permutation.
\end{theoreme} 
\begin{proof}
\label{sympol} As it was shown in Lemma \ref{5.1} the first terms in asymptotic expansion of $c_n$ are of the following form:
\begin{gather*}
\frac{\sum_{M_i}\chi(M_i)}{6}+\frac{N}{4}, \quad n=4;\\
\sum_{i=1}^N \frac{2k-1}{4\lambda _{i+N,i+N}^{2k+1}},\quad n=2k+1, \,k>1;\\
\sum_{M_i}\frac{a_{ki}\Gamma(k+1)}{4\pi}+\sum^{N}_{i=1}  \frac{2k}{4\lambda _{i+N,i+N}^{2k-2}}, \quad n=2k+2, \,k>1.
\end{gather*}

Consider $x_i=\frac{1}{\lambda_{i+N,i+N}}$, $i=1,\cdots, N$. Taking the first $N$ coefficients of powers of $z$ starting from $z^5$ and denoting them by $d_{n+4},\, n=1, \cdots, N$ we obtain $N$ equations which have the following structure
\begin{gather*}
d_{n+4}=\sum_i x_i^{n}
\end{gather*}
To find the solution of such a system we find
the following symmetric polynomials in terms of $c_n$ using a 
well-known  recursive procedure:
\begin{gather*}
S_1=x_1+\dots+x_{N},\\
S_2=x_1x_2+x_1x_3+\dots+x_2x_3+\dots,\\
S_3=x_1x_2x_3+x_1x_2x_4+\dots+x_2x_3x_4+\dots,\\
\cdots\\
S_{N}=x_1x_2\dots x_{N}.
\end{gather*}
Due to the Vi\`ete Theorem we state now that $x_1, \dots x_{N}$ are the
roots of the equation
\begin{gather*}
x^{N}-S_1x^{N-1}+S_2x^{N-2}+\dots+S_{N}=0
\end{gather*}
and can be found up to permutation because we assume that they are mutually distinct.

If we substitute the obtained values of $\lambda_{i+N,i+N}$ 
in the expression obtained in Lemma \ref{5.2}  we  obtain a system  of linear equations for $|\lambda_{i,i+N}|$ and can find them.

As soon as we have found these elements, the result of Lemma \ref{5.3}  gives us  a system  of linear equations for  $\lambda_{i,i}$ and we can also find them.
The Theorem is proved.
\end{proof}

\end{document}